\documentclass[aps,pra,twocolumn,amsmath,amssymb,nofootinbib,showpacs,superscriptaddress]{revtex4-1}

\usepackage[caption=false, position=top, font=footnotesize]{subfig}
\usepackage[english]{babel}
\usepackage{latexsym}
\usepackage{graphics}
\usepackage{graphicx, tabularx}
\usepackage{epsfig}
\usepackage{color}
\usepackage{bm}
\usepackage{amsmath}
\usepackage{amssymb}
\usepackage{amsthm}
\usepackage{dcolumn}
\usepackage{bm}
\usepackage{float}
\usepackage{hyperref}
\usepackage{color}
\usepackage{epstopdf}
\usepackage[svgnames]{xcolor}
\usepackage{cleveref, braket, comment,bbold, adjustbox, tikz}
\hypersetup{hidelinks,colorlinks=true,allcolors=blue}
\usepackage[normalem]{ulem}
\usetikzlibrary{quantikz2}

\renewcommand{\r}{\ensuremath{{\sf R}}}
\newcommand{\cnx}{\ensuremath{{{\sf C}^{N-1}\sf{X}}}}
\newcommand{\cnz}{\ensuremath{{{\sf C}^{N-1}\sf{Z}}}}

\renewcommand{\r}[3]{R_{#1}^{#2}(#3)}

\newcommand{\ph}[2]{{\sf Ph}^{#1}(#2)}
\newcommand{\xx}[2]{{\sf XX}^{#1}(#2)}

\newcommand{\czn}[1]{{\sf C}^{#1}{\sf Z}}
\renewcommand{\r}[3]{R_{#1}^{#2}(#3)}

\newcommand{\cxn}[1]{{\sf C}^{#1}{\sf X}}

\begin{document}

\preprint{APS/123-QED}

\title{Universal quantum computing with qubits embedded in trapped-ion qudits}

\author{Anastasiia S. Nikolaeva}
\affiliation{Russian Quantum Center, Skolkovo, Moscow 121205, Russia}
\affiliation{National University of Science and Technology ``MISIS”,  Moscow 119049, Russia}
\author{Evgeniy O. Kiktenko}
\affiliation{Russian Quantum Center, Skolkovo, Moscow 121205, Russia}
\affiliation{National University of Science and Technology ``MISIS”,  Moscow 119049, Russia}
\author{Aleksey K. Fedorov}
\affiliation{Russian Quantum Center, Skolkovo, Moscow 121205, Russia}
\affiliation{National University of Science and Technology ``MISIS”,  Moscow 119049, Russia}

\date{\today}
\begin{abstract}
Recent developments in qudit-based quantum computing, in particular with trapped ions, open interesting possibilities for scaling quantum processors without increasing the number of physical information carriers.
Here we propose a method for compiling quantum circuits in the case, where qubits are embedded into qudits of experimentally relevant dimensionalities, $d=3,\ldots,8$, for the trapped-ion platform. 
In particular, we demonstrate how single-qubit, two-qubit, and multiqubit gates can be realized using single-qudit operations and the M$\o$lmer-S$\o$rensen (MS) gate as a basic two-particle operation.
We expect that our findings are directly applicable to the current generation of trapped-ion-based qudit processors.
\end{abstract}

\maketitle

\section{Introduction.}
Despite remarkable progress in the development of semiconductor microelectronics~\cite{Moore1965}, a number of problems remain practically intractable for modern computing architectures,
e.g., prime factorization for large numbers~\cite{Shor1994} and simulation of complex (quantum interacting, many-body) systems~\cite{Feynman1982}.
Quantum computing, which is based on the idea of manipulating entangled superposition states of qubits~\cite{Deutsch1985,Brassard1998,Ladd2010} (quantum counterparts of classical bits), 
is considered as a way to extend computational capabilities in solving hard computational problems including aforementioned tasks~\cite{Shor1994,Lloyd1996,Fedorov2022}.
A serious challenge is, however, to develop a large-scale quantum computing device that allows solving practically relevant tasks.
Specifically, it is nontrivial to design a physical platform that admits scalability to large number of qubits without degrading the quality of the control for their interactions.
At the current frontier, the task is to build noisy intermediate-scale quantum (NISQ) devices~\cite{Bharti2021}, where qubits are noisy and operations between them are imperfect. 
Recent progress in developing NISQ devices includes the work on various physical platforms such as
superconductors~\cite{Gambetta2018,Martinis2019,Pan2021-4}, semiconductors~\cite{Vandersypen2022,Morello2022,Tarucha2022}, 
neutral atoms~\cite{Lukin2021,Browaeys2021,Browaeys2020-2,Saffman2022}, 
trapped ions~\cite{Monroe2017,Blatt2012,Blatt2018},
and quantum optical systems~\cite{Pan2020,Lavoie2022}.

The used information carriers within aforementioned platforms are essentially multilevel, so their use as qubits require idealization. 
A natural approach is then to consider such systems as {\it qudits}, which are $d$-dimensional quantum systems ($d>2$).
Quantum computing with qudits has been activity studied during last decades~\cite{Farhi1998,Kessel2002,Muthukrishnan2000,Nielsen2002,Berry2002,Klimov2003,Bagan2003,Vlasov2003,Clark2004,Leary2006,Straupe2006,Straupe20062,Ralph2007,White2008,Straupe2008,Ionicioiu2009,Ivanov2012,
Kiktenko2015,Kiktenko2015-2,Song2016,Bocharov2017,Gokhale2019,Pan2019,Low2020,Martinis2009,White2009,Straupe2010,Wallraff2012,Mischuck2012,Gustavsson2015,Martinis2014,Ustinov2015,Morandotti2017,Balestro2017,Low2020,Sawant2020,Senko2020,Pavlidis2021,Rambow2021,OBrien2022,Nikolaeva2022, lopez2019qubit}. 
Notably, the first realization of two-qubit gates has used qudits~\cite{Wineland1995}: two qubits were stored in the degrees of freedom of a single trapped ion.
Recent results include demonstrations of multiqubit processors based on trapped ions~\cite{Ringbauer2021,Kolachevsky2022}, superconducting~\cite{Hill2021,Schuster2022}, and optical systems~\cite{OBrien2022}.
In the case of trapped ions~\cite{Ringbauer2021,Kolachevsky2022} efficient experimental control of systems with up to eight levels with high-enough gate fidelities has been shown,
and experimental results for systems with even higher level numbers have been presented~\cite{Low2023}.

On the top of overcoming various experimental challenges, the development of qudit-based processors also requires new approaches for realizing quantum algorithms and, specifically, compiling quantum circuits having qudits as a resource.
Already for the case of qutrits $(d=3)$, additional level is useful for decompositions of multiqubit gates~\cite{Ralph2007,White2009,Ionicioiu2009,Wallraff2012,Gokhale2019,Nikolaeva2022}.
In addition, one can consider qudit’s space as a space of multiple qubits, e.g. ququart ($d=4$) as two qubits~\cite{Kiktenko2015-2,Kiktenko2015,Kolachevsky2022}.
For larger qudit dimensionalities, one can combine these approaches~\cite{Nikolaeva2021}. 
In general, the development of resource-efficient compilation schemes for various dimensionalities on the basis of native operations remains a problem, 
which is crucial for realizing quantum algorithms~\cite{Nikolaeva2021,Mato2022,Ringbauer2023, Fischer2023}. 
Using a trapped ion platform as an example, we investigate how the multilevel resource of qudits allows solving it elegantly.

In this work, we demonstrate how universal quantum computation can be realized with the use of multilevel quantum systems potential. 
For this purpose, we develop a method for compilation of qubit quantum circuits on trapped-ion qudits with various dimensionalities. 
We show how to assist the third level of qutrits $(d=3)$ to make a generalized Toffoli gate for qubits located in the subspace of the first two levels.
Next, we present a way how to obtain a universal gate set for a system of qubits in pairs embedded in ququarts $(d=4)$.
We then extend this consideration to ququints ($d=5$), quxehes ($d=6$), and qusepts ($d=7$) by showing how additional levels can be used for decomposition of multiqubit gates.
Finally, we consider the universal gate set for qubits' triples embedded in quocts ($d=8$).
As our approach uses the MS gate, we expect that our results are directly applicable for the current generation of trapped-ion-based qudit processors and cover all the relevant dimensionalities. 

\section{Model of a trapped-ion qudit-based processor }

We first define a model of a trapped-ion qudit-based processor.
We consider a system of $m$ qudits with equal number of available levels $d$, whose basis states are denoted by $\ket{j}$, $j=0,\ldots,d-1$.
As a basic set of available single-qudit operations we employ Pauli rotations
$\r{\phi}{ij}{\theta}=\exp(-\imath\sigma^{ij}_{\phi}\theta/2)$,
where $\theta$ is a rotation angle, $i\neq j$ denotes two addressed levels of the qudit, $\sigma_\phi^{ij} = \cos\phi \sigma_x^{ij} + \sin\phi \sigma_y^{ij}$, 
$\sigma^{ij}_\kappa$ with $\kappa = x,y,z$ stands for a standard Pauli matrix acting in a two-level subspace spanned by $\ket{i}$ and $\ket{j}$ (e.g., $\sigma_y^{02}=-\imath\ket{0}\!\bra{2}+\imath\ket{2}\!\bra{0}$), and angle $\phi$ specifies a rotation axis.
We fix notations for rotations around $x$- and $y$-axis:
$\r{x}{ij}{\theta}:= \r{\phi=0}{ij}{\theta}$, $\r{y}{ij}{\theta}:= \r{\phi=\pi/2}{ij}{\theta}$.
We also assume the possibility to implement a phase gate $\ph{j}{\theta}$, which adds a phase factor $e^{\imath\theta}$ to the basis state $\ket{j}$ and keeps other basis states unchanged.
We note that making a single-qudit rotation at any desired transition is possible by combining rotations among a connected graph of physically allowed transitions~\cite{Mato2022}. On the technical side, qudits are controlled by a single laser acousto-optic modulator.

As a basic two-qudit operations, we consider a phase-compensated qudit-based extensions of the MS gate $\widetilde{\sf MS}_{\phi}^{ijkl} (\chi) := \exp(-\imath  \sigma^{ij}_{\phi}\otimes \sigma^{k\ell}_{\phi}\chi)$,
where $\chi$ describes the duration of the gate, and pairs $(i,j)$, $(k,l)$ indicate addressed transitions in coupled qudits~\cite{Ringbauer2021}. 
This operation can be obtained from the phase-compensated MS-gate
$
    \widetilde{\sf MS}_{\phi}^{01}(\chi)=\exp\left[-\imath (\sigma_\phi^{01}\otimes\sigma_\phi^{01})\chi\right]
$
by surrounding it with single-qudit gates (see App. \ref{app:a}).
Phase-compensating procedure of the standard MS-gate
$
  {\sf MS}_{\phi}^{01}(\chi)=\exp\left[-\imath (\sigma_\phi^{01}\otimes \mathbb{1}+ \mathbb{1}\otimes \sigma_\phi^{01})^2\chi/2\right]
$
is presented in Ref.~\cite{Ringbauer2021}.
We also fix notations for $x$-axis-based rotations: ${\sf XX}^{ijkl}({\chi}) := \widetilde{\sf MS}_{\phi=0}^{ijkl}(\chi)$.
An ability to perform two-qudit gates is assumed for every pair of qudits, that is a natural assumption for trapped ions~\cite{Ringbauer2021,Kolachevsky2022}.

We also assume an ability to perform a computational basis read-out measurement, described by positive operator valued-measure (POVM)
${\cal M}=\{\ket{j}\bra{j}\}_{j=0,\ldots,d-1}$. 
It can be realized physically by repeating projective quantum non-demolition (QND) measurement $\{\ket{0}\bra{0}, \sum_{j\geq 1}\ket{j}\bra{j}\}$ interleaved with a proper population exchanging between qudit's levels.
This allows us to make a computational basis measurement of all kinds of embeddings of qubits in qudits considered further.

\begin{figure}[]
\centering
\includegraphics[width=\linewidth]{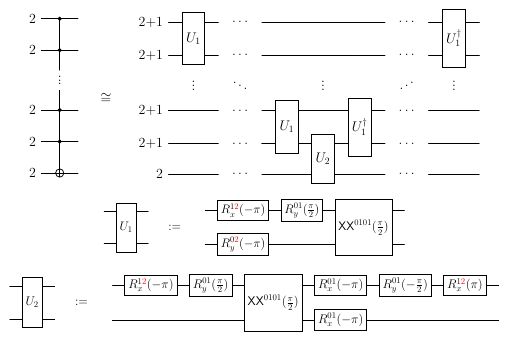}
\caption{Decomposition of the generalized Toffoli gate for qubits embedded in trapped-ion qutrits.
The number near each wire specifies the employed dimensionality of the corresponding particle.
The sign $\cong$ emphasizes that the circuit in the RHS is equivalent to the one in the LHS for qubits' subspace occupying the first two levels only.}
\label{fig:qutrit_toff}
\end{figure}

\section{Realizing circuits with trapped-ion-based qudits of various dimensions}

Below, we consider realizing qubit-based circuits by embedding qubits in trapped-ion qudits of experimentally relevant range of dimensions from $d = 3$\footnote{Preliminary \href{https://conference.rqc.ru/session/3}{result} on qutrit-based decomposition was presented as a poster 
	“Qudit-based quantum compiler: realizing the full potential of multi-level quantum systems”, A.S. Nikolaeva, E.O. Kiktenko, A.K. Fedorov,  at the VI 
 International conference on quantum technologies "ICQT2021". An extended version of the presented result was \href{https://mipt.ru/priority2030/info/64\%20\%D0\%BD\%D0\%B0\%D1\%83\%D1\%87\%20\%D0\%BA\%D0\%BE\%D0\%BD\%D1\%84\%20\%D0\%9B\%D0\%A4\%D0\%98_1.pdf}{published} in the proceedings of the 64th All-Russian Scientific Conference of MIPT (talk "Generalized Toffoli gate decomposition on superconducting and trapped-ion-based qutrits", A.S. Nikolaeva, E.O. Kiktenko, A.K. Fedorov).} to $d=8$.

\subsection{Simplifying decomposition of multiqubit gates with qutrits}

The computational space of qutrit ${\sf Q}$ can be considered as a space of qubit ${\sf q}$ ($\ket{0}_{\sf Q} \leftrightarrow \ket{0}_{\sf q},  \ket{1}_{\sf Q} \leftrightarrow \ket{1}_{\sf q}$) together with an ancillary level $\ket{a=2}_{\sf Q}$.
In this way, $m$ qubits are embedded in $m$ qutrits.
This embedding is actually an extension of a standard qubit-based approach, therefore to obtain a universal gate set for qubits' subspaces, it is enough to employ $\r{\phi}{01}{\theta}$ and $\xx{0101}{\chi}$ gates without populating third levels.

However, the ancillary level $\ket{a=2}$ appears to be quite practical for implementing multiqubit generalized Toffoli (multi-controlled NOT) gate
\begin{multline}
    \cnx: \ket{c_1,\ldots,c_{N-1},t}
    \\
    \mapsto \ket{c_1,\ldots,c_{N-1}, t\oplus {\prod}_{i=1}^{N-1} c_i},
\end{multline}
where $c_i,t\in\{0,1\}$ and $\oplus$ stands for mod 2 summation. 
In Fig.~\ref{fig:qutrit_toff}, we illustrate a V-shape ladder-like decomposition of the \cnx~gate, which assists upper ancillary levels of qutrits.
The shown circuit realizes \cnx~operation within qubits' subspace of involved qutrits in a way that all ancillary levels remain unpopulated by the end of the circuit, given that they were unpopulated at the start.

The presented decomposition requires $2N-3$ two-qutrit $\xx{0101}{\pi/2}$ gates applied within the nearest-neighbor linear coupling map.
To compare, known qubit-based decompositions of multi-controlled gates require ${\cal O}(N^2)$ two-qubit gates without using ancillary qubits~\cite{Barenco1995}.
We emphasize that the decomposition in Fig.~\ref{fig:qutrit_toff} is based on trapped-ions-specific two-qutrit MS gate,
which is different (even after being surrounded by local gates) to controlled phase and iSWAP two-qutrit operations, considered previously~\cite{Wallraff2012, Kiktenko2020, Nikolaeva2022} for simplifying the realization of generalized Toffoli gates.
It also differs from the decomposition based on the Cirac-Zoller gate~\cite{fang2023}. 
In contrast to the MS gate, the Cirac-Zoller gate requires preparing the motional degree of freedom in the ground state.

\subsection{Embedding pairs of qubits in ququarts}

The feature of ququarts, compared to qutrits, is that their space can be considered as a space of two qubits.
Namely, two qubits, ${\sf q}$ and ${\sf q}'$, can be embedded into ququart ${\sf Q}$ using the mapping $\ket{j}_{\sf Q} \leftrightarrow \ket{{\rm bin}_2(j)}_{{\sf q}{\sf q}'}$, where ${\rm bin}_k(j)$ stands for $k$-bit binary representation of $j=0,\ldots,2^k-1$ (e.g., ${\rm bin}_2(0)=00$, ${\rm bin}_2(1)=01$, etc.)
This approach allows storing $n$ qubits in $\lceil n/2 \rceil$ ququarts.

\begin{figure}
\centering
\includegraphics[width=\linewidth]{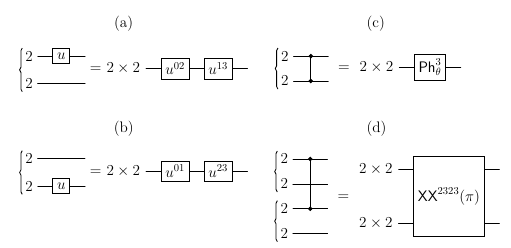}
\caption{Compilation of single- and two-qubit gates for qubits embedded in trapped-ion-based ququarts. 
Braces indicate qubits embedded in the same qudit.
Realizations of single-qubit gate $u$, acting on the first and second qubit inside a qudit, are shown in (a) and (b) respectively.
Here, superindices $ij$ indicate levels at which ququart gate $u^{ij}$ operates.
In (c), a realization of a two-qubit controlled-phase gate acting on qubits inside the same qudit is shown.
In (d), an implementation of a two-qubit controlled-phased gate between qubits from different ququarts using the MS gate is depicted.
}
\label{fig:ququart_gates}
\end{figure}

Since each qubit ${\sf q}_1$ is always located together with some `neighboring' qubit ${\sf q}_2$ inside a ququart ${\sf Q}$, a single qubit gate $u$, acting on ${\sf q}_1$, has to be considered as a tensor product with the identity operator, acting on ${\sf q}_2$.
Depending on a `position' of ${\sf q}_1$ inside ${\sf Q}$, the single-qudit unitary $U$ that implements $u$ takes the form:
\begin{equation}
    \begin{aligned}
    U &= u\otimes\mathbb{1} = 
    \begin{pmatrix}
        u_{00} & & u_{01} & \\
        & u_{00} & & u_{01} \\
        u_{10} & & u_{11} & \\
        & u_{10} & & u_{11} \\
    \end{pmatrix},\\
    U &= \mathbb{1}\otimes u=
    \begin{pmatrix}
        u_{00} & u_{01} & & \\
        u_{10} & u_{11} & & \\
        & & u_{00} & u_{01}\\
        & & u_{10} & u_{11}\\
    \end{pmatrix},
    \end{aligned}\label{eq:sqqrt-from-sqb}
\end{equation}
where unspecified elements are zeros and $u_{ij}=\bra{i}u\ket{j}$ [see also Fig.~\ref{fig:ququart_gates}(a, b)].
Note that the number of independent parameters defining ququart unitary $U$ in Eq.~\eqref{eq:sqqrt-from-sqb} is the same as for a single-qubit unitary $u$.
Given single-qubit gate $u=\r{\phi}{}{\theta}=: \r{\phi}{01}{\theta}$, the ququart-based realization, depending on the affected qubit, takes the form
$\r{\phi}{}{\theta}\otimes\mathbb{1} = \r{\phi}{02}{\theta}\r{\phi}{13}{\theta}$ or
$\mathbb{1}\otimes \r{\phi}{}{\theta} = \r{\phi}{01}{\theta}\r{\phi}{23}{\theta}$.
Similarly, single-qubit gate $u=\ph{}{\theta}=:\ph{1}{\theta}$ can be realized by $\ph{}{\theta}\otimes\mathbb{1}=\ph{2}{\theta}\ph{3}{\theta}$ or by $\mathbb{1}\otimes\ph{}{\theta}=\ph{1}{\theta}\ph{3}{\theta}$.

The realization of a two-qubit gate, e.g.,
${\sf CZ}: \ket{xy}_{{\sf q}_1{\sf q}_2}\mapsto (-1)^{xy}\ket{xy}_{{\sf q}_1{\sf q}_2}$,
depends on whether ${\sf q}_1$ and ${\sf q}_2$ belong to the same qudit ${\sf Q}$ or not.
In the former case, it can be realized with the single-ququart gate ${\sf CZ} = \ph{3}{\pi}= {\rm diag}(1,1,1,-1)$
[see also Fig.~\ref{fig:ququart_gates}(c)].
In the latter case, we have to consider a 16-dimensional four-qubit (two-ququart) operation given by a proper tensor product of ${\sf CZ}$ with identity matrices.
It appears that $\xx{2323}{\pi}$ acting on ququarts pair ${\sf Q}_1$, ${\sf Q}_2$  adds $-1$ phase factor to four basis states $\ket{1a1b}_{{\sf q}_{11}{\sf q}_{12}{\sf q}_{21}{\sf q}_{22}}$ with $a,b\in\{0,1\}$, where ${\sf q}_{ij}$ 
denotes $j$th qubit embedded in ququart ${\sf Q}_i$.
This operation is exactly ${\sf CZ}$ gate [see also Fig.~\ref{fig:ququart_gates}(d)].
Realizations of ${{\sf CZ}}$ gates on qubit pairs $({\sf q}_{11}, {\sf q}_{22})$, $({\sf q}_{12}, {\sf q}_{21})$, and $({\sf q}_{12}, {\sf q}_{22})$ are provided by $\xx{2313}{\pi}$, $\xx{1323}{\pi}$, and $\xx{1313}{\pi}$, respectively.

Having an ability to apply $\r{\phi}{}{\theta}$ and $\ph{}{\theta}$ operations to arbitrary qubits, and two-qubit ${\sf CZ}$ to an arbitrary pair of qubits, we obtain access to the universal set of qubit gates.
The developed approach decreases the number of required two-particle operations required for running a given circuit compared to the straightforward qubit-based realization.
This is due to the fact that some two-qubit operations act on qubits in the same qudits.
In particular, a standard ancilla-free decomposition of ${\sf C}^3{\sf X}$ gate, which requires $14$ two-qubit gates~\cite{nakanishi2021quantumgate},
maps to the one with $6$ two-ququart gates; see App. \ref{app:b}.

\begin{figure}
\centering

\includegraphics[width=\linewidth]{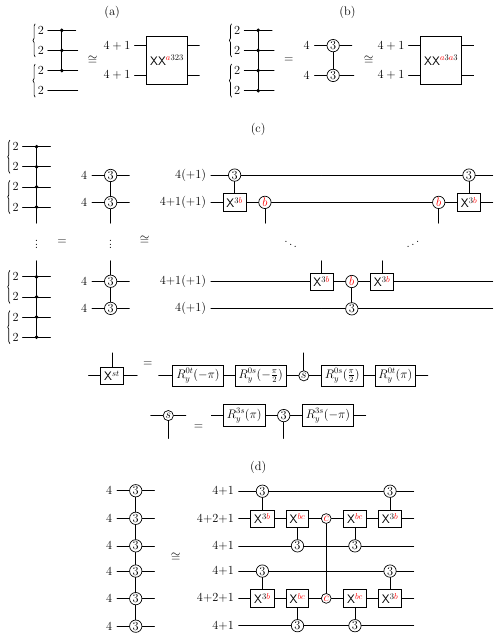}
\vspace{-15pt}
\caption{{Realizing $\czn{N-1}$ gate for qubit pairs embedded in qudits with $d>4$ using assistance of upper levels.
Vertical lines connecting circled 3s denote multiqudit gates that add a phase -1 if and only if all affected qudits are in the state $\ket{3}$.
Note that this operation is equivalent to controlled-phase gate acting on all embedded qubits.
In (a) and (b), the implementations of  ${\sf C}^2{\sf Z}$ and ${\sf C}^3{\sf Z}$ with the help of an additional level $a$ and a single $\xx{ijk\ell}{\pi}$ gate are shown correspondingly.
Note that the implementation of ${\sf C}^3{\sf Z}$ makes a ${\sf Ph}^{33}_{{\cal A}_{\overline{a}}}$ operation.
In (c), a linear depth decomposition of multiqubit $\czn{N-1}$ gate based on two-qudit ${\sf Ph}^{33}_{{\cal A}_b}$ gates is shown.
These gates can be realized either by adapting straightforward qubits' decomposition of $\czn{3}$ gate (see App. \ref{app:b}) or by employing an extra ancillary level $a\neq b$, that is marked by (+1), and using scheme (b).
In (d) log-depth decomposition of the $\czn{11}$ gate by using the ${\sf Ph}^{33}_{{\cal A}_{b,c}}$ gates, which employs two ancillary levels $b$ and $c$, is shown. 
These gates can be realized using the third ancillary level $a\neq b,c$ and scheme (b).}
\vspace{-12pt}}
\label{fig:ququint_4q_tof}
\end{figure}

\subsection{Improving decompositions of multiqubit gates with 5, 6, and 7th levels.}
The $d$-level qudit space ($d=5,6,7$) can be considered as a $4$-dimensional subspace of two qubits directly summed with a $(d{-}4)$-dimensional subspace of ancillary levels.
This allows using all the techniques developed previously for ququarts, with additional benefit from operating with ancillary levels.

As shown in Fig.~\ref{fig:ququint_4q_tof}(a, b) an additional ancillary level $a$ (e.g., $a=4$ for $d=5$) reduces the required number of two-particle operations for making ${\sf C}^2{\sf Z}$ and ${\sf C}^3{\sf Z}$ qubit gates 
down to a single implementation of $\xx{a323}{\pi}$ and $\xx{a3a3}{\pi}$, correspondingly.
Importantly, the implementation of ${\sf C}^3{\sf Z}$ gate makes the transformation 
${\sf Ph}^{33}_{{\cal A}_{\overline{a}}}$, where
\begin{equation}
    {\sf Ph}^{33}_{\cal A}:
    \ket{j,k} \mapsto 
    \begin{cases}
        (-1)^{\delta_{3,j}\delta_{3,k}}\ket{j,k} 
            &\text{~for~}(j,k)\in {\cal A},\\
        e^{\imath \phi(j,k)}\ket{j,k}
            &\text{~for~}(j,k)\not\in {\cal A},\\
    \end{cases}
\end{equation}
$\delta_{i,j}$ stands for standard Kronecker symbol, $\phi(j,k)$ are some real phases, and ${\cal A}_{\overline{a}}:=\{(x,y): x,y\in\{0,\ldots,d-1\}/\{a\}\}$.
In other words, ${\sf C}^3{\sf Z}$ is a matrix of the form ${\rm diag}(1,\ldots,1,-1,1,\ldots,1)$ in the two-qudit subspace spanned by all levels except $a$.

At the same time, as shown in Fig.~\ref{fig:ququint_4q_tof}(c), $\cnz$ gate acting on $N$ qubits ($N$ is even) embedded in $N/2$ qudits can be implemented with only $N-3$ two-qudit operations ${\sf Ph}^{33}_{{\cal A}_b}$ with ${\cal A}_b:=\{(x,y): x\in\{0,\ldots,3,b\}, y\in\{0,\ldots,3\}\}$.
${\sf Ph}^{33}_{{\cal A}_b}$ can be implemented in two ways.
The first option is to adopt a qubit-based decomposition of the four-qubit ${\sf C}^3{\sf Z}$ gates (see App. \ref{app:b}).
It does not employ any additional ancillary levels and can be realized for $d=5$ qudits.
In this case, each of ${\sf Ph}^{33}_{{\cal A}_b}$ operations requires 6 two-qudit $\xx{ijkl}{\pi}$ gates, and the whole decomposition of $\cnz$ requires $6N-18$ native two-qudit gates.
We also note that the sequence of gates in the second ``ascending''
part of the circuit have to be taken as a Hermitian conjugate to the first ``descending'' part.
The second option to realize ${\sf Ph}^{33}_{{\cal A}_b}$ is to make a single ${\sf Ph}^{33}_{{\cal A}_{\overline{a}}}$ operation with $a\neq b$ (note that ${\cal A}_{\overline{a}}\supset {\cal A}_b$).
It is possible for $d\geq 6$.
In the result, we obtain a decomposition of $\cnx$ gate with $N-3$ operations $\xx{a3a3}{\pi}$.

Moving from $d=6$ to $d=7$ allows an additional improvement in multiqubit gates decomposition.
In particular, as shown in Fig.~\ref{fig:ququint_4q_tof}(d), the $\cnx$ gate can be realized with logarithmic depth circuit consisting of $N-3$ two-ancillary-levels-based~${\sf Ph}^{33}_{{\cal A}_{b,c}}$ operation, where ${\cal A}_{b,c}:=\{(x,y): x,y\in\{0,\ldots,3,b,c\}\}$.
Again, ${\sf Ph}^{33}_{{\cal A}_{b,c}}$ can be reduced to a single $\xx{a3a3}{\pi}$ operation with $a\neq b,c$.

We summarize the properties of decompositions in Table~\ref{tab:table1}.
One can see that the more ancillary levels are under control, the easier the multiqubit gate decomposition is.

\begin{table}[b]
	\caption{Properties of \cnx~gate decompositions for $N$ qubits embedded in $N/2$ qudits ($N$ is even).}
	\begin{ruledtabular}
		\begin{tabular}{c|c|c}
			{Qudit dimension} & {Number of $\xx{ijk\ell}{\chi}$ gates} & {Depth} \\
			\hline
			$d=5$ & $12N-18$ & $\mathcal{O}(N)$ \\
			$d=6$ & $N-3$ & $\mathcal{O}(N)$ \\
			$d=7$ & $N-3$ & $\mathcal{O}(\log N)$ \\
		\end{tabular}
	\end{ruledtabular}\label{tab:table1}
\end{table}

\begin{table}[b]
	\caption{Number of entangling gates in Grover's algorithm realization with qubits, qutrits, and ququints.}
  \centering
	\begin{ruledtabular}
 \centering
		\begin{tabular}{>{\centering\arraybackslash}m{1.8cm}|>{\centering\arraybackslash}m{2.9cm}|>{\centering\arraybackslash}m{3.0cm}}
			{Num. of qubits in the algorithm} & Num. of ${\sf CZ}$ gates in the qubit-based transpilation & Num. of ${\sf XX}(\chi)$ gates in the qudit-based transpilation \\
			\hline
                $n=3$ & 7 & 4 (for $d=3$)\\
                $n=4$ & 40 & 4 (for $d=5$)\\
		\end{tabular}
	\end{ruledtabular}\label{tab:table2}
\end{table}

\begin{figure}
\centering
\includegraphics[width=\linewidth]{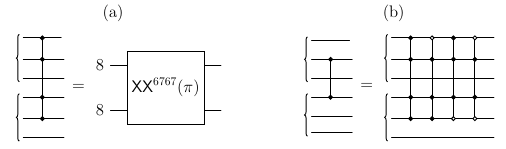}
    \vspace{-20pt}
    \caption{Implementing gates that act on qubits embedded in two different quocts.
    In (a), the realization of ${\sf C}^3{\sf Z}$ with single $\xx{6767}{\pi}$ two-qudit gate is shown. 
    In (b), the implementation of ${\sf CZ}$ gate between two qubits from different qudits with 4 ${\sf C}^3{\sf Z}$ gates is presented.
    Standard notation for inversed controls (empty dots) is used.
    The inversion can be realized by surrounding the basic ${\sf C}^3{\sf Z}$ gate with $R_x(\pm \pi)$ single qubit gates extended to single qudit operations.}
    \label{fig:quocts}
    \vspace{-10pt}
\end{figure}

We also note that the controlled phase gate ${\sf Ph}^{33}_{{\cal A}}$ with ${\cal A}=\{(x,y): x,y\in\{0,\ldots,3,a_1,a_2,\ldots\}\}$ for ancillary levels $a_1,a_2,\ldots$ is the basic one for a number of other decompositions considered extensively in Ref.~\cite{Nikolaeva2021}.
Any of the decompositions for $d$-level qudits developed in Ref.~\cite{Nikolaeva2021} can be realized with $(d+1)$-level trapped-ion qudits using the scheme from Fig.~\ref{fig:ququint_4q_tof}(b).

\subsection{Embedding triples of qubits in quocts}

The ability to control $d=8$ levels allows embedding three qubits ${\sf q}_1, {\sf q}_2, {\sf q}_3$ in a single qudit ${\sf Q}$, e.g., by a straightforward mapping $\ket{j}_{\sf Q}\leftrightarrow \ket{{\rm bin}_3(j)}_{{\sf q}_1,{\sf q}_2,{\sf q}_3}$.
Single-qubit operations as well as gates between qubits inside the same qudit can be realized by single-qudit operations.
To obtain a universal qubit-based gate set, we consider the $\xx{ijk\ell}{\pi}$ operation for certain $i,j,k,\ell$.
It can be used to realize a ${\sf C}^3{\sf Z}$ gate between two pairs of qubits from different qudits [see an example in Fig.~\ref{fig:quocts}(a)]. 
The resulting gate set is universal, as can be seen from the fact that a ${\sf CZ}$ gate between any qubits from different qudits can be obtained by making four ${\sf C}^3{\sf Z}$ gates, as shown in Fig.~\ref{fig:quocts}(b).

\section{Discussion}
In this work, we have proposed a set of methods for implementing qubit circuits using trapped-ion qudits of experimentally relevant dimensions $d=3,\dots,8$.
To the best of our knowledge, this is the first time that it is explicitly shown how to realize a universal qubit gate set, consisting of a single-qubit and entangling ${\sf CZ}$ gate, acting on qubits embedded in pairs (in qudits with $d\geq 4$) or in triples (in qudits with  $d\geq 8$).
Efficient ancilla-free realizations of generalized Toffoli gates by employing extra levels of trapped ion-based qudits have also been demonstrated.
The resulting number of required entangling gates for making the ${\sf C}^{N-1}{\sf X}$ gate is given by $c_1 N + c_2$ (where $c_i$ are some constants), and $c_1$ decreases with qudit dimension $d$.
We note that all our decompositions employ the MS gate, which is native for trapped ion-based architecture.
These features distinguish the current contribution from other recent proposals on qudit-based decompositions mostly based on controlled phase gates (see, e.g,~\cite{White2009, Wallraff2012, Gokhale2019, Kiktenko2020, Nikolaeva2021}).

\begin{figure*}
\centering
\includegraphics[width=\linewidth]{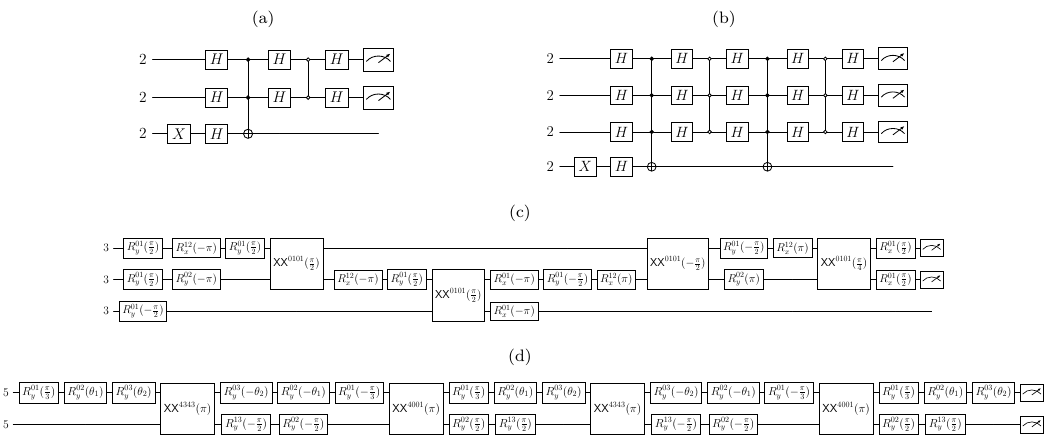}
    \caption{Circuits of Grover's algorithm implementation.
    In (a) and (b) hardware-agnostic circuits for searching $``11"$ and $``111"$ hidden strings with $2+1$ and $3+1$ qubits are presented correspondingly. 
    Empty circles correspond to controls on state $\ket{0}$.
    In (c) circuit from (a) is transpiled to the single-qutrit and two-qutrit ion gates. 
    Multiqubit decomposition from Fig.~\ref{fig:qutrit_toff} is used.
The circuit depicted in (d) corresponds to the ququint-based realization of circuit (b). 
    Straightforward qubit-to-ququint mapping is used: the first (second) qubit pairs are embedded in the first (second) ququint. The highest levels of ququints are used to simplify the implementation of multi-qubit gates, as presented in Fig.~\ref{fig:ququint_4q_tof} (a,b).
    Here, $\theta_1 := 2\arcsin(3^{-1/2})
    $, $\theta_2 := \pi/2$.
    }
    \label{fig:grover}
    \vspace{-10pt}
\end{figure*}

A striking example of obtaining an advantage in realizing a concrete algorithm with our approach (compared to a straightforward qubit-based version) is Grover's algorithm~\cite{Grover1997}. 
In Fig.~\ref{fig:grover}(a,b) we provide high-level qubit circuits for Grover's algorithm with oracles of dimensions $n=3$ and $n=4$ correspondingly (``hidden string'' as selected as $(n-1)$ units).
According to~\cite{Barenco1995,nakanishi2021quantumgate}, ancilla-free qubit-based decomposition of ${\sf C}^{N-1}{\sf X}$ (${\sf C}^{N-1}{\sf X}$) for $N=3$ and $N=4$ gates requires 6 and 14 ${\sf CZ}$ entangling gates correspondingly.
This results in 7 and 40 entangling gates for realizing $n=3$ and $n=4$-qubit Grover's algorithm correspondingly.
At the same time, qudit-based transpilation makes it possible to reduce the number of entangling operations down to 4 as shown in Fig.~\ref{fig:grover}(c,d).
Here we provide schemes for qudits of dimensions $d=3$ and $d=5$ as demonstrative examples.
The results of comparison are also shown in Table~\ref{tab:table2}.

An important question is whether an advantage brought by qudit-assisted reduction of the gate number prevails over the extra noise and consequential errors coming from employing upper qudit levels.
To address this question, we provide a simplistic consideration that allows one to estimate a necessary relation between errors in qubit- and qudit-based schemes to assure the advantage of the latter.
Let us assume that the main source of errors is represented by entangling gates, and $e_{\rm b(d)}$ is the resulting error probability in a qubit (qudit) entangling operation that appeared due to the noise in the system.
Let $N_{\rm b(d)}$ be the number of entangling gates in a qubit (qudit)-based circuit realizing some fixed operation, e.g., a generalized Toffoli gate.
The probability of at least one error in the qubit (qudit) case reads
\begin{equation}
    E_{\rm b(d)} = 1 - (1-e_{\rm b(d)})^{N_{\rm b(d)}} \approx N_{\rm b(d)} e_{\rm b(d)},
\end{equation}
given that $e_{\rm b(d)}$ is small enough.
Then one has $E_{\rm d}< E_{\rm b}$ in the case of
\begin{equation}
    \frac{e_{\rm d}}{e_{\rm b}} < \frac{N_{\rm b}}{N_{\rm d}} =:
    \kappa.
\end{equation}
For the developed schemes, $\kappa \rightarrow 3$ [Fig.~\ref{fig:qutrit_toff}(a)], $\kappa\rightarrow 6$ [Fig.~\ref{fig:qutrit_toff}(b,d)], or even $\kappa = \infty$ [Fig.~\ref{fig:ququart_gates}(c)] (here the limit is taken with respect to the qubit number $N$ in the decomposition of the ${\sf C}^{N-1}{\sf Z}$ gate).
Taking into account the progress in suppressing errors on the upper qudit levels with continuous decoupling~\cite{decoupling2023} and the potential improvement of error suppression with magnetic shielding~\cite{Ringbauer2023}, we expect the developed qudit-based techniques to bring a real advantage compared to qubit-based realizations.
We also refer the reader to recent work~\cite{jankoviс2023noisy}, where a question related to an influence of qudit and qubit errors, yet considered in an another context, is analyzed.

\section{Conclusion}

We have presented a set of techniques for compilation of quantum circuits on qubits that are embedded in trapped-ion qudits. 
A feature of the proposed approach is the use of qudit-based generalization of 
MS gate as a basic two-particle gate.
We have shown how increasing the dimensionality in experimentally relevant cases (from $d=3$ up to $d=8$) brings more and more benefits in the compilation of qubit quantum circuits.
We expect these results are directly applicable to trapped-ion qudit-based processors~\cite{Ringbauer2021,Kolachevsky2022}.

\section*{Acknowledgements}

We thank B.I. Bantysh and I.V. Zalivako for fruitful discussions.
The research is supported by the Priority 2030 program at the NIST ``MISIS'' under the project K1-2022-027.
The work of A.S.N. and E.O.K. was also supported by the RSF Grant No. 19-71-10091 (development of qutrit-based and ququart-based decompositions).

\appendix
\renewcommand{\r}[3]{R_{#1}^{#2}(#3)} 

\section{Extending qubit trapped-ion gates to qudit space}\label{app:a}

\begin{figure}
    \centering
\includegraphics[width=0.95\linewidth]{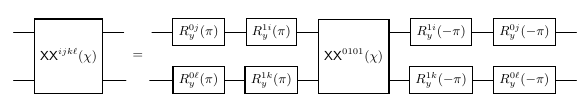}
\caption{$\xx{ijk\ell}{\chi}$ gate implementation on the basis of $\xx{0101}{\chi}$ gate with single-qudit rotations.}
\label{fig:xijkl}
\vspace{-15pt}
\end{figure}

In Fig.~\ref{fig:xijkl}, we show how to transform a $\xx{0101}{\chi}$ gate, usually considered a basic two-qudit operation in trapped-ion qudit-based processors~\cite{ Ringbauer2021,Kolachevsky2022}, to a more general $\xx{ijk\ell}{\chi}$ gate.
We note that each transition of the form $\r{\phi}{ij}{\theta}$ can be decomposed to a sequence of operations within a connected coupling map between $d$ levels of a qudit.
As an illustrative example, we consider $^{171}{\rm Yb}^+$-based ququarts ($d=4$)~\cite{Kolachevsky2022} with the coupling map consisting of four transitions: $\ket{0}\leftrightarrow \ket{1}$, $\ket{0}\leftrightarrow \ket{2}$, $\ket{0}\leftrightarrow \ket{3}$.
The transition on an arbitrary level pair $(i,j)$ takes the form
\begin{equation}
	\r{\phi}{ij}{\theta} = 
	\r{y}{0i}{\pi}
	\r{\phi}{0j}{\theta}
	\r{y}{0i}{-\pi}.
\end{equation}

\begin{figure*}
    \centering
 \includegraphics[width=\linewidth]{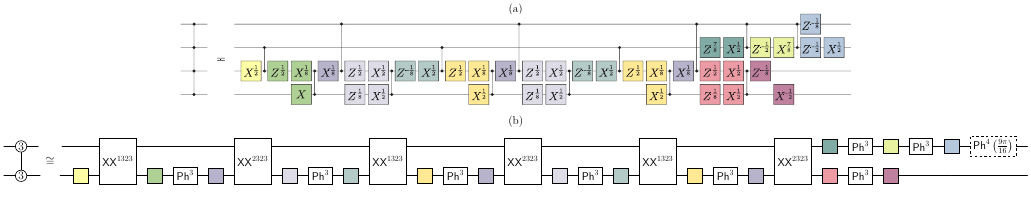}
    \caption{In (a), a qubit-based decomposition of ${\sf C}^3{\sf Z}$ gate from Ref.~\cite{nakanishi2021quantumgate} is shown.
    Here, ${X}^{\alpha}:=R_x(\alpha\pi)$ and ${Z}^{\alpha}:=R_z(\alpha\pi)$.
    In (b), the corresponding implementation for four qubit embedded in two qudits is shown.
    Here, ${\sf XX}^{ijk\ell}:=\xx{ijk\ell}{\pi}$ and ${\sf Ph}:=\ph{3}{\pi}$.
    Empty squares of different colors denote sequences of single-qudit gates that are extensions of corresponding (sequences of) single-qubit gates of the same color.
    Two-qubit ${\sf CZ}$ gates between qubits embedded in the same qudits are shown as ${\sf Ph}^3$ gates explicitly.
    The highlighted single-qudit gates in the end of the circuit makes a phase correction, which necessary for ancillary-level-based decompositions.
    }
    \label{fig:c3z-ququarts}
\end{figure*}

\section{Adapting qubit-based decomposition of $\cxn{3}$ gate for four qubits embedded in two qudits}\label{app:b}

Here, we show how to implement $\czn{3}$ (equivalently, $\cxn{3}$) gate with $\xx{ijk\ell}{\chi}$ operation for four qubits embedded in two qudits without employing  auxiliary levels.
For this purpose, we employ the best to our knowledge no-ancilla qubit-based decomposition from Ref.\cite{nakanishi2021quantumgate}, shown in Fig.~\ref{fig:c3z-ququarts}(a).
It consists of 14 two-qubit ${\sf CZ}$ gates within all-to-all coupling map and a number of single-qubit gates.
Mapping this scheme to two qudits results in a circuit with 6 two-qudit $\xx{ijk\ell}{\pi}$ gates as shown in Fig.~\ref{fig:c3z-ququarts}(b).
Each of $\xx{ijk\ell}{\pi}$ gates corresponds to one of ${\sf CZ}$ gate.
Remaining $14-6=8$ two-qubit ${\sf CZ}$ gates transform into single-qudit $\ph{3}{\pi}$ gates, since the corresponding qubits belong to the same qudits.
All single qubit $2\times 2$ unitaries maps to single-qudit $d\times d$ ones according to tensor product and tensor sum (in the case of $d>4$) rules.

To use this decomposition as a basic block for other decompositions assisted by upper levels (for $d=5$), one needs to take into account an appearing phase difference between the operation on the first four qubit levels and upper levels.
This phase difference appears due to the cross terms in tensor products of single-qudit operators of the form $(u\otimes \mathbb{1}_{\rm qb}) \oplus \mathbb{1}_{\rm anc}$ (or $(\mathbb{1}_{\rm qb} \otimes u) \oplus \mathbb{1}_{\rm anc}$), 
where $u$ is a single-qubit operation $\mathbb{1}_{\rm qb}$ and $\mathbb{1}_{\rm anc}$ are identity operators acting in the space of the neighboring qubit and the space of upper levels, correspondingly.
In particular, to employ the decomposition from Fig.~\ref{fig:c3z-ququarts}(a) to make ${\sf Ph}^{33}_{{\cal A}_4}$ gate, which transforms $\ket{33}$ to $-\ket{33}$ and keeps $\ket{xy}$ unchanged for $x\in\{0,\ldots,4\}, y\in\{0,\ldots,3\}, xy \neq 33$, 
it is required to make an additional phase correction operation in the end of the circuit in Fig.~\ref{fig:c3z-ququarts}(b).

\newpage

\bibliography{bibliography-qudits.bib}

\end{document}